\documentclass[pre,aps,twocolumn,superscriptaddress,amsmath,showpacs,floatfix]{revtex4}
\usepackage{graphicx}
\begin{document}

\title{Anomalous Dynamics of Forced Translocation}
\author{Yacov Kantor}
\email{kantor@post.tau.ac.il}
\affiliation{School for Physics and
Astronomy, Raymond and Beverly Sackler Faculty of Exact Sciences,
Tel Aviv University, Tel Aviv 69978, Israel}
\author{Mehran Kardar}
\affiliation{Department of Physics, Massachusetts Institute of
Technology, Cambridge, Massachusetts 02139}

\date{\today}

\begin{abstract}
We consider the passage of long polymers of length $N$ through a hole in a membrane. 
If the process is slow, it is in principle possible to focus on the dynamics of the
number of monomers $s$ on one side of the membrane, assuming that the two segments
are in equilibrium.
The dynamics of $s(t)$ in such a limit would be diffusive, with a mean translocation
time scaling as $N^2$ in the absence of a force, and   proportional to $N$
when a force is applied. We demonstrate that the assumption of equilibrium
must break down for sufficiently long polymers (more easily when forced),
and provide lower bounds for the translocation time by comparison to unimpeded
motion of the polymer.
These lower bounds exceed the time scales calculated on the basis of equilibrium,
and point to anomalous (sub--diffusive) character of translocation dynamics. 
This is explicitly verified by numerical simulations of the unforced translocation
of a self-avoiding polymer.
Forced translocation times are shown to strongly depend on the method by which
the force is applied. In particular, pulling the polymer by the end leads to much
longer times than when a chemical potential difference is applied across the membrane.
The bounds in these cases grow as $N^2$ and $N^{1+\nu}$, respectively, where
$\nu$ is the exponent that relates the scaling of the radius of gyration to $N$.
Our simulations demonstrate that the actual translocation times scale in the same
manner as the bounds, although influenced by strong finite size effects which persist
even for the longest polymers that we considered ($N=512$). 
\end{abstract}
\widetext

\maketitle

\section{Introduction}

Translocation of a polymer  through a narrow pore in a membrane
is important to many biological processes,
such as the injection of viral
DNA into a host, DNA packing into a shell during viral
replication, and gene swapping through bacterial pili
\cite{MolBioCell}. Translocation also has practical applications
in genetics as in cell transformation by DNA electroporation
\cite{MolBioCell}, and in gene therapy \cite{gt}. 
This has inspired a number of recent {\em in
vitro} experiments, including the electric field-induced
migration of DNA through microfabricated channels \cite{han}, or
through an $\alpha$-hemolysin protein channel in a membrane
\cite{Kasianowicz,meller}. 
Experiments are motivated by the
possibility to ``read-off" a DNA or RNA sequence by tracking its  
passage through a pore \cite{Kasianowicz,meller,kam}.

Translocation of a polymer involves both molecular considerations, such
as the shape of the pore channel and its interactions with DNA,
as well as more macroscopic factors such as the statistics and dynamics
of the long polymer.
It is the universal features of the latter which are the focus of this  
paper.
While worming its way  through the hole, the segments of the polymer on
each side of the membrane can ``explore" many possible configurations.
The number of allowed configurations actually is least when the polymer
is halfway through the hole, presenting an entropic barrier.
In this regard translocation  resembles other entropically controlled  
polymer
systems, such as polymer trapping in random environments
\cite{entrap1,entrap2,entrap3}, DNA gel
electrophoresis \cite{electrophor} or reptation
\cite{deGennes_book}, where the geometry of the obstacles
constrains the kinetics of the polymer.

\begin{figure}
\includegraphics[width=7cm]{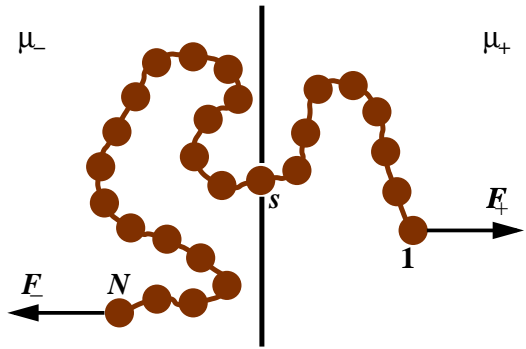}
\caption{\label{transloc} Schematic depiction of  polymer
translocation  from the left side of a membrane to its right side.
(Throughout the paper we shall follow this
convention for  the direction of transport.)  The number $s$ of the monomer at the  
hole is denoted the ``translocation coordinate.''}
\end{figure}

A number of recent theoretical works have shed light on the
translocation process
\cite{Kasianowicz,muthu,muth2,deGennesDNA,Lubensky,bs,zandi},
which is schematically depicted in Fig.~\ref{transloc}.
A single variable $s$ representing the
monomer number at the pore \cite{muthu,Lubensky,park} indicates
how far the polymer has progressed, and is
a natural variable for describing this problem.
Due to its resemblance to the
`reaction coordinate' for chemical processes, we call $s$ the
translocation coordinate. If $s$ changes very slowly, such
that the polymer segments on both sides of the membrane
have time to equilibrate, the mean force acting on the monomer in the
hole can be determined from a simple calculation of free energy, and
the translocation problem is then reduced to the escape of a
`particle' (the translocation coordinate) over a potential barrier.
In the following, we shall refer to this limit as {\em Brownian  
translocation},
but shall demonstrate that the required equilibration is not tenable for
long enough polymers.

In many experimental situations the polymers are not very long and
the observed behavior strongly depends on the properties of the
polymer and the pore. Thus introduction of some specific
features into the channel properties \cite{Lubensky,slonkina}, or into
interparticle potential \cite{loebl,kong}, may provide qualitative
explanations of the observed behavior. While the theoretical
understanding of the experiments is growing, we are still quite
far from quantitative understanding of many observed features
\cite{meller_review}. In this work, we restrict out attention to
qualitative features of very long polymers that are independent
of the details of the pore or inter-monomer potentials.
Consequently, we restrict ourselves to simple models.
Furthermore, since we are interested in contrasting  the behavior of
phantom polymers, in which the monomers do not interact with 
each other, with self--avoiding ones, in which monomers repel
each other at short distances, we perform simulations in 
two--dimensional (2D) space, where such differences are more
pronounced.

The reduction of the motion of a large molecule to a
single--particle problem ignores the fact that the positions of the
monomers have strong correlations \cite{deGennes_book}, leading to
non-trivial dynamical effects \cite{deGennes_book,Doi}. In
particular, in a dilute solution of polymers in a good solvent, on
time scales shorter than the overall relaxation time of the
polymer, the motion of monomers is characterized by {\em
anomalous} dynamics \cite{KBG,carm}. Not surprisingly, such
effects are also present in translocation. In
a previous work we demonstrated \cite{ckk} that the scaling of
the translocation times with the number of monomers has a
power--law dependence which cannot be derived from Brownian
motion of a particle  over a barrier, but rather follows
from general scaling considerations. In this work, we focus on
 consequences of anomalous dynamics
in the presence of a force.

The rest of this paper is organized as follows.
In Sec.~\ref{sec_brown} we review the limit of very slow translocation,
in which case the problem can be reduced to the Brownian motion
of a single coordinate.
We demonstrate that the requirement of equilibration breaks down
for long polymers, especially in the presence of
a force pulling the polymer to one side.
We also demonstrate the importance of how such a force is applied to
the polymer, contrasting the cases of a polymer pulled by the end,
with one forced into a favorable environment.
Lower bounds on the translocation times are obtained in Sec.~\ref{sec_unimpeded}
by comparison with the unimpeded motion of a polymer.
The long time scale for equilibration of a forced polymer is due to its change
of shape, e.g.  into a stretched sequence of blobs when pulled at one end.
Exactly how this change of shape is achieved through transmission of the
force from one end to another is explicitly shown in Appendix~\ref{app_oneD}
which solves this problem for a one--dimensional (1D) phantom polymer.
Our other model system, the self-avoiding two dimensional polymer is
discussed in Appendix~\ref{app_twoD}.
The bounds from unimpeded motion serve a convenient reference point for discussion
 of anomalous processes that are described in detail in Sec.~\ref{sec_anomal}. 
In particular, we find that the actual translocation times have the same
scaling behavior as the lower bounds.

\section{Brownian translocation and its limitations}\label{sec_brown}

If the translocation process is sufficiently slow, at each stage the
statistics of the segments will be governed by the equilibrium
Boltzmann weight.
If so, the dynamics is constrained to reproduce the corresponding  
statistics.
A simple method to achieve this is to focus on the translocation
coordinate $s$, and to write down a stochastic Langevin equation for its
evolution.
This procedure, which we shall refer to as {\em Brownian translocation},
is the chief tool employed in most analytical  
studies\cite{muthu,Lubensky,park},
and shall be reviewed in this Section.

\subsection{Unforced motion}

In the absence of an external potential, the entropic contributions to
the free energy of the two polymeric segments result in a
free energy ${\cal F}(s)=\gamma k_BT\ln[(N-s)s]$ \cite{muthu,Lubensky,park}.
Note that there is a decrease in the number of possible states as
the polymer threads the hole, which can be regarded as an entropic  
barrier.
For phantom polymers (random walks)  $\gamma$ is equal
to 1/2, while for self-avoiding polymers it depends on the dimensionality of  
space.
 From this free energy, we can construct a Langevin equation
$\dot s=-m {\partial{\cal F}/\partial s}+\eta(t)$, where $m$ is a
mobility coefficient indicating how easily the polymer is pulled  
through
the hole. To obtain the correct Boltzmann statistics, the noise  
$\eta(t)$ has
to be uncorrelated at different times, with a variance equal to $m  
k_{B}T$.

As demonstrated in Ref.~\cite{ckk}, the corresponding Fokker--Planck
equation for the probability $p(s,t)$
can be made {\em independent} of $N$ by a simple change of variables
$s\rightarrow sN$ and $t\rightarrow t/N^2$. Consequently, the average
translocation time $\tau$ (and its fluctuations)
must scale as $N^2$ for any $\gamma$.
In fact, a very similar distribution for the transit time is obtained by
ignoring the potential barrier completely (setting $\gamma=0$).
In this limit the translocation coordinate simply undergoes diffusion,
i.e. at time scales much shorter than typical translocation times
we expect $\left\langle \left(s(t')-s(t)\right)^{2}\right\rangle\propto  
|t-t'|$.

\subsection{Pulling on the ends}
The advent of optical tweezers has made it possible to manipulate  
single macromolecules.
A common procedure is to attach latex balls to the end(s) of a polymer  
(such as DNA),
and then to manipulate the balls by optical tweezers \cite{gerland,farkas}.
While this is not the method commonly used in the translocation  
experiments,
it motivates an interesting extension of the previous calculation.
Let us imagine that through an optical tweezer set-up forces  $F_{+}$  
and $F_{-}$
are applied to the two ends of the polymer (perpendicular to the wall),
as in Fig.~\ref{transloc}.

A polymer configuration in which the ends are separated by a distance  
$\vec r$
gets an additional Boltzmann weight of $\exp\left(\vec F\cdot\vec  
r/k_{B}T\right)$.
For a Gaussian polymer of length $N$, integration over all locations  
$\vec r$ leads to
a contribution of $\exp\left[N(Fa/k_{B}T)^2\right]$ to the partition  
function
($a$ is a microscopic length scale).
Restricting the integration over $\vec r$ only to half space (as  
appropriate
for the translocation problem), does not change the qualitative form,
i.e., in the presence of the force the free energy acquires a term
proportional to $k_{B}T N (Fa/k_{B}T)^2$.

The above argument can be generalized to a self-avoiding polymer by  
noting
that due to dimensional considerations a force $F$ should always  
appear in
the combination $FR/k_{B}T$, where $R$ is a characteristic size of the  
polymer, such as its radius of gyration $R_g$.
Quite generally $R\sim a N^{\nu}$, with $\nu={1}/{2}$ for a {\em  
phantom} 
(non--self--interacting) polymer, and $\nu={3}/{4}$ and 0.59
for self--avoiding polymers in space dimensions $d=2$ and 3,
respectively \cite{madras}.
The contribution of the force to the free energy can thus be written as
$k_{B}T\Phi\left(FaN^{\nu}/k_{B}T\right)$.
When the force is sufficiently strong to deform the polymer into a  
sequence of
blobs (we shall argue later that this is the relevant regime for  
translocation),
the free energy is expected to be linear in $N$, necessitating  
$\Phi(x)\sim x^{1/\nu}$.
Such considerations thus imply a free energy contribution of $N k_{B}T
\left(Fa/k_{B}T\right)^{1/\nu}$.
In the specific case of translocation, adding up the contributions from  
the two segments
leads to
\begin{equation}\label{F(f,s)}
{\cal F}(s)\sim k_{B}T\left[s\left(F_{+}a\over k_{B}T\right)^{1/\nu}+
(N-s)\left(F_{-}a\over k_{B}T\right)^{1/\nu}\right].
\end{equation}

The corresponding Langevin equation for the translocation coordinate is  
now
\begin{equation}\label{dots(f)}
\dot s = \lambda\left(F_{+}^{1/\nu}-F_{-}^{1/\nu}\right)+\eta(t),
\end{equation}
where we have absorbed various coefficients into the parameter  
$\lambda$.
Note that the average velocity has a non-linear dependence on the  
forces;
in the case of $F_{-}=0$ growing as $F_{+}^{1/\nu}$.
Consequently, the translocation time in such a set up should decrease  
with the
applied force as $\tau\propto N/F_{+}^{1/\nu}$.
Note that this expression breaks down for {\em small forces}, where the  
typical translocation
time is controlled by the diffusive fluctuations.
The distinction between weak and strong force regimes has a specific  
meaning in the
context of polymers, and quantified through the scaling combination  
$\tilde{f}\equiv FaN^{\nu}/k_{B}T$.
For weak forces this combination is small, and the equilibrium polymer  
shape is not changed.
For strong forces $\tilde{f}\gg 1$, and the polymer becomes  
stretched.
The same division applies to forces that are strong enough to overcome  
the diffusive
character of the translocation coordinate.

\subsection{Chemical potential difference}
A more common situation for translocation is when the
environments separated by the membrane are not equivalent, so that
the polymer encounters a chemical potential difference between the two
sides. In this case, the leading contribution to the free energy is
${\cal F}(s)=\mu_{+}s+(N-s)\mu_{-}$, and the Langevin equation
takes the form
\begin{equation}\label{dots(mu)}
\dot s = \lambda' \Delta\mu+\eta(t),
\end{equation}
with $\Delta\mu\equiv\left(\mu_{-}-\mu_{+}\right)>0$.
In this case the average translocation velocity is predicted to be  
proportional
to $\Delta\mu$, leading to typical exit times that scale as  
$N/\Delta\mu$
for large $\Delta\mu$.
It is possible to envision situations in which the polymer is forced
(or hindered) by a combination of both a chemical potential
difference, and forces applied to the two ends.
For the corresponding Langevin equation, we merely need to
add the force contributions in Eqs.~(\ref{dots(f)}) and (\ref{dots(mu)}).

In the experiments of Meller {\em et al.}~\cite{meller}, translocation 
is driven by an
electric potential difference between the two sides of an artificial
membrane suspended in a liquid.
It is commonly assumed that since the
conductivity of the liquid is significantly higher than the
conductivity of the membrane, the liquid on each side of the
membrane is an equipotential.
The voltage drop then occurs only across the membrane,
and  is experienced only by the
(charged) monomers moving through the pore \cite{meller_review}.
If so, the voltage difference is equivalent to the chemical potential
difference discussed above.
However, the previous results with force acting only on the end monomers
serve as a warning that the results are quite sensitive to where the  
force
is applied to the polymer.
It would thus be reassuring to carry out a more precise calculation of
the electric field in the vicinity of the pore, and how it acts upon  
the monomers.

\subsection{Limitations}
The analytical procedure outlined in this section rests on the  
assumption
that the two polymeric segments have come to equilibrium,
so that the corresponding free energy can be used to construct a
Langevin equation.
The minimal requirement is that the typical translocation time
$\tau$ should exceed the equilibration time 
$\tau_{\rm equil}$ of a polymer.
For a chain of finite size, it is always possible to achieve this by
designing the pore to have a large friction coefficient.
In Ref.~\cite{Lubensky}, it is argued that this is the case applicable 
to the experiments of Ref.~\cite{meller}. However, the equilibration 
time of a polymer depends strongly on its length, scaling as 
$\tau_{\rm equil}(N)\propto N^{z\nu}$. As discussed in the next 
section, the exponent $z\nu$ is typically  larger than 2 for 
Brownian dynamics of self-avoiding polymers (and equal to 2 for a 
phantom polymer).

Since typical transit times for unforced translocation scale as $N^{2}$,
it is possible to imagine that the formalism may apply to phantom  
polymers. Indeed there is some evidence of this from numerical  
simulations \cite{chern},
although with an inexplicably large friction coefficient.
The situation becomes worse in the presence of a force (either imposed
by pulling or a chemical potential difference), in which case typical
translocation times are predicted to be proportional to $N$.
In the latter case, the range of applicability of `Brownian  
translocation'
is even further limited.

\section{Unimpeded motion of a polymer}\label{sec_unimpeded}

Since the collective dynamics of the passage of polymer through the pore
is hard to treat analytically, as a first step we shall derive
{\em lower bounds} on the characteristic time scale.
The key observation is that the restriction that the monomers should
sequentially pass through a hole in a membrane can only impede the motion
of the polymer. Hence the time scale for the polymer to travel the
same distance in the absence of the wall should be a generous
lower bound to its translocation time.
In this Section we shall thus explore the time scale for unimpeded
motion of the polymer in the circumstances of interest.

\subsection{Unforced diffusion}

In the unforced limit, the translocating polymer simply goes from one
side of the membrane to the other by  `diffusion.'
In the process, the center of mass of the polymer moves a distance of  
the
order of a typical size, say the gyration radius of $R_g$.
How long does it take for a polymer to move a similar distance
without the constraints imposed by the pore and the wall?
In the absence of hydrodynamic interactions, the diffusion constant $D$  
of a polymer
is related to the diffusion constant $D_0$ of a single monomer by  
$D=D_0/N$.
Consequently, the time that a polymer needs to diffuse its own
radius of gyration scales as $N^{1+2\nu}$ \cite{deGennes_book}.
(This is also the relaxation time of the slowest internal mode of
a polymer \cite{deGennes_book}, and is called the Rouse equilibration
time.)
For self--avoiding polymers $\nu>1/2$, and the equilibration time
is clearly longer than that obtained for translocation though a hole
in the wall assuming Brownian translocation.
The Rouse time scale of $N^{1+2\nu}$ should thus be a lower bound
to the correct translocation time.

\subsection{Pulling on the end}

Now consider a polymer that is being pulled by one end.
The regime of interest to us is when the force is strong enough to
deform the shape of the polymer.
The equilibrium shape of the polymer is then a stretched 
sequence of `blobs' \cite{deGennes_book}.
The number of monomers per blob $N_B$ is such that force acting
on it is marginally strong, and obtained from $FaN_B^\nu\sim k_BT$.
The typical size of each blob is thus $R_B\sim aN_B^\nu\sim k_BT/F$,
while the number of blobs is $N/N_B\sim N(Fa/k_BT)^{1/\nu}$.
The overall length of the stretched chain is now
$R(F)\sim R_B(N/N_B)\sim aN(Fa/k_BT)^{1/\nu-1}$.
The mobility of the center of the mass of a polymer of length $N$ is
proportional to $1/N$, and since there is only a force applied to one
monomer, its net velocity scales as $F/N$.
The characteristic relaxation time is the same as the time scale of
the polymer moving a distance of order of its size, and hence behaves
as 
\begin{equation}\label{tau(f)}
\tau_{\rm equil}(F)\sim\frac{R(F)}{v(F)}\propto N^2 F^{-2+1/\nu}.
\end{equation}
(The same conclusion is obtained if we start with a globular polymer
and then apply the force to one end, and wait until the other end
feels the force.)

Note that upon approaching the boundary between weak and strong forces
at $F\propto N^{-\nu}$, we regain the equilibration time $N^{1+2\nu}$
for unforced polymers. 
However, the result in Eq.~(\ref{tau(f)}) is only valid for
$N^{-\nu}\leq Fa/k_BT\leq 1$, since for stronger forces, as we shall
see in the specific models described in the next Section, the velocity
of the monomer (and hence of the entire chain) saturates.

Unimpeded motion of the pulled polymer thus places a lower bound
of $N^2$ on the translocation time, far exceeding the time scale
($\propto N$) calculated in the previous section.
It may not be readily apparent how the force applied to one end
of the polymer is transmitted to the other end, and why the
qualitative picture of blobs presented above is valid.
There is actually one limit in which the problem of pulling a polymer
by the end can be solved analytically, and that is for a 1D
phantom polymer. In Appendix \ref{app_oneD} this model and the 
corresponding analysis are presented in some detail.

\subsection{Mimicking a chemical potential difference}

It is difficult to come up with an unhindered situation that best
resembles the case of a chemical potential difference across a membrane.
Absent the restrictions imposed by the membrane, there is now a force
that is applied to a single monomer, at the spatial position where the
membrane would reside. 
Unlike the previous case, the monomer to which the force is applied
now changes constantly. 
There is thus no incentive for a drastic change in the shape of polymer,
and we assume that the scaling of the size remains the same,
i.e. $R\sim a N^\nu$, independent of $\Delta\mu$.
At each moment there is a force of $\Delta \mu/a$ applied to the
entire polymer, as a consequence of which its center of mass 
should move with a velocity $v\propto \Delta\mu/N$.
We thus conclude that the time for such unhindered motion
scales as
\begin{equation}\label{tau(mu)}
\tau(\Delta\mu)\sim\frac{R}{v}\propto \frac{N^{1+\nu}}{\Delta\mu}.
\end{equation}

Note that to recover the equilibration time of the unforced polymer we
have to set $\Delta\mu\propto N^{-\nu}$ in the above equation. 
While this is the same scaling form as that of a force applied to the end,
it is different from the weak/strong criterion that would have been
deduced on the basis of energetics ($\Delta\mu N\sim k_BT$).
This is a reflection of the manner in which we introduced the unimpeded
version (as a force, rather  than a chemical potential difference.)
Nonetheless, we still expect the velocity, and hence the time scale
in Eq.~(\ref{tau(mu)}) to saturate for $\Delta\mu\sim k_BT$,
as explicitly demonstrated for the models considered in the next Section.

\section{Anomalous translocation}\label{sec_anomal}

Having established some some (presumably generous) lower bounds,
we now would like to focus on the true asymptotic dynamics of translocation.
Given the limitations of analytical studies, the chief tool employed in
this section is numerical simulations. 
Interestingly, we find that the lower bounds obtained in the previous
section are actually quite restrictive.

\subsection{Sub--diffusive behavior of unforced motion}

In a previous work \cite{ckk}, we made a detailed study of the $N$--dependence
of the mean translocation time.
One of the central conclusions was that in the case of Brownian dynamics
of a self-avoiding polymer, the translocation time scales as
\begin{equation}\label{tau}
\tau\sim N^{1+2\nu}.
\end{equation}
This is of the same order as the equilibration time of a polymer of length $N$,
and also demonstrates that the actual exit time scales in the same manner
as the bound established in the previous section.

The above $N$--dependence of $\tau$ is inconsistent with simple 
diffusion of the translocation coordinate $s(t)$,
reflecting the constraints imposed by the collective motion of the
entire polymer. 
A related situation occurs for the fluctuations of a labeled monomer in space,
which are also anomalous and sub--diffusive\cite{KBG} on time scales shorter
than the equilibration time.
Following this analogy, we suggested \cite{ckk} that the short time fluctuations
of $s(t)$ follow the anomalous diffusion relation
\begin{equation}\label{andif}
\langle\Delta s^2
(t)\rangle\sim t^{2\zeta}.
\end{equation}
For Eqs.~(\ref{tau}) and (\ref{andif}) to be consistent, we must obtain
$\Delta s$ of order $N$, when $t$ of order $\tau$,
leading to the exponent relation $\zeta=1/(1+2\nu)$. 
Note that for a phantom polymer $\nu=\zeta={1}/{2}$,
i.e. the anomaly disappears in this limit, and the process becomes diffusive.
(This differs from the corresponding motion of a labeled
monomer \cite{KBG,carm}, which remains anomalous
even for a phantom polymer.)
This is consistent with a detailed study of a three--dimensional phantom polymer by 
Chern {\em et al.}~\cite{chern} which concluded that the results may be interpreted
in terms of diffusive motion of the translocation coordinate over a barrier. 
Such correspondence is likely a fortuitous coincidence for phantom
polymers, and even in this case, the value of the effective
diffusion constant could not be obtained from the
geometrical features of the model \cite{chern}.

\begin{figure}
\includegraphics[width=8cm]{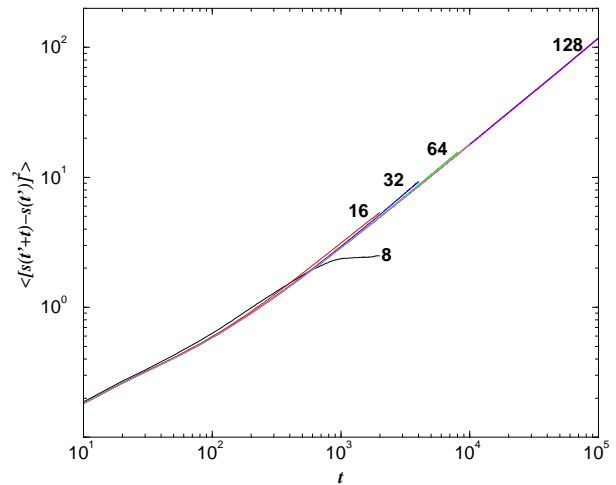}
\caption{\label{cor_lglg_no_force} 
Temporal fluctuations of the monomer number $s$ located at the hole,
averaged over the initial time $t'$ and over 1000 independent
simulations, for polymers of lengths $N=8$, 16, 32, 64, and 128. }
\end{figure}

To establish the anomalous nature of (unforced) translocation
dynamics, we carried out Monte Carlo (MC) simulations on a model of
{\em two-dimensional self-avoiding} polymers, described in Appendix~\ref{app_twoD}.
Simulations in two (rather than three) dimensions have the advantage of relative 
ease, and stronger differences from phantom polymers.
We followed the dynamics of $s(t)$, focusing on the quantity
$\Delta s^2\equiv (s(t'+t)-s(t'))^2$.
This correlation function is depicted in Fig.~\ref{cor_lglg_no_force},
and was obtained by averaging over $t'$ of over 1000 independent
simulations  for $N=8$, 16, 32, 64 and 128. 
Two cautionary points must be made in considering this data:
The first is that we have no a priori assurance that this process is stationary;
the results may depend on both $t$ and $t'$, and consequently
influenced by the averaging over $t'$.
(For example, choosing a short averaging range may
increase the effect of the initial conditions. We performed
averaging over shorter ranges of $t'$ and did not see significant
differences in the correlation functions.)
Secondly, for each value of $t$, we can only include processes whose translocation
time exceeds $t$; consequently, the size of the ensemble decreases with
increasing $t$. (However, this effect is insignificant for  $t$
several times shorter than the mean translocation time.) 
In Fig.~\ref{cor_lglg_no_force} all available $t'$ are included, i.e.
for each polymer the statistics was collected up to the moment
that the translocation was completed.

Since the values of $s$ cannot exceed $N$, as the time--difference
$t$ becomes of order of $\tau$, $\Delta s^2$
saturates, as is apparent in the case of $N=8$ in Fig.~\ref{cor_lglg_no_force}. 
However, for times shorter than $\tau(N)$, the results for different
lengths seem to form a single curve. 
This instills confidence in the quasi--stationary character of translocation on time
scales shorter than $\tau$. On the logarithmic scale the graph
seems to have curvature for $t<1000$. This is probably a consequence
of discreteness of the model, since corresponding
differences in $s$ are smaller than 5. For larger times the slope
of the curve approaches 0.80, which clearly indicates the presence
of anomalous diffusion, and is consistent with the expected value
$2/(1+2\nu)$, with $\nu={3}/{4}$ for 2D self--avoiding walks \cite{madras}. 
Thus, despite its finite duration, the translocation process at short scales
resembles a stationary process (at least $\Delta s^2$ is insensitive to $t'$)
which exhibits anomalous dynamics.

\begin{figure}
\includegraphics[width=8cm]{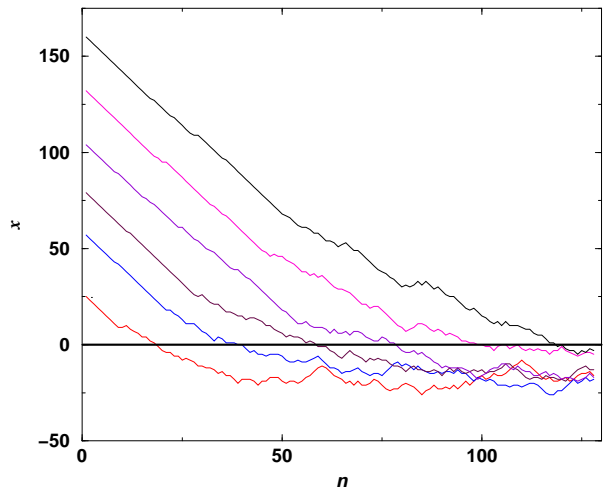}
\caption{\label{phant_pic} ``Snapshots'' of a 128-monomer  phantom
polymer passing through a membrane in one dimension. Each line depicts
the position $x$ (in lattice units) of the $n$th monomer at a fixed time.
Different lines correspond to times when the 20th, 40th, \dots 120th (bottom-left to top-right)
monomer crosses the membrane (the thick line at $x=0$).
}
\end{figure}

\subsection{Pulling on the end}

As discussed in Sec.~\ref{sec_brown}, the polymer pulled by a force $F_{+}$
is relatively undistorted as long as $\tilde{f}\equiv F_+a N^\nu/k_BT\ll 1$,
and the corresponding translocation times are not very different to those
in the absence of force.
Increasing polymer length at fixed $F_+$ ultimately leads to a regime
with $\tilde{f}\gg 1$, in which the polymer is expected to be stretched into
a sequence of blobs. 
It is the latter regime which is of interest us, and which shall be explored
by examining the one-- and two--dimensional polymer models introduced earlier.

\subsubsection{One--dimensional phantom polymer}

Simulations are carried out with the model 1D phantom
polymer presented in Appendix~\ref{app_oneD},
starting with a polymer that is equilibrated on one
(say, left) side of the ``membrane'' (a point on a
1D lattice) with one end point held at the membrane. 
The ``narrow opening" is implemented by allowing only {\em sequential passage}
of the monomers across the membrane, i.e. the
$n$th monomer can move from the left to the
right only if $(n-1)$th monomer is already on the right side. 
Conversely, the $n$th monomer may diffuse from the right side to the left, only
if the $(n+1)$th monomer is on the left side. 
The first monomer of the chain is restricted to remain on the
right throughout the process.
We study the dynamics of translocation as a function of the 
force $F$ applied to the first monomer.
However, the results become independent of $F$ when the reduced force
$f\equiv Fa/k_BT$ exceeds unity, since it becomes very unlikely for 
the first monomer to move backwards.

\begin{figure}
\includegraphics[width=8cm]{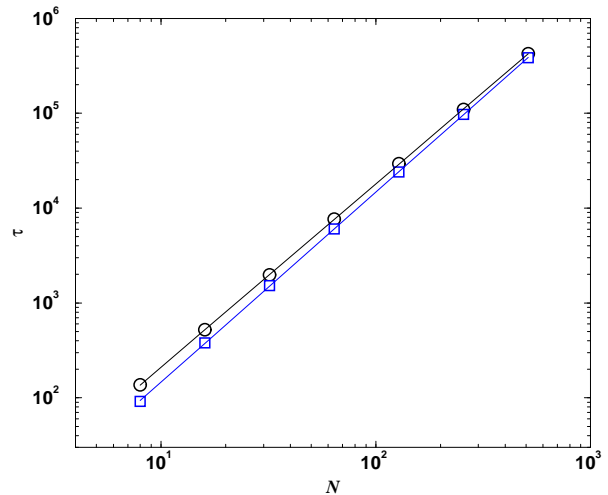}
\caption{\label{inf_F_phant} Logarithmic plot of the mean
translocation time as a function of length for a
one--dimensional phantom polymer with an infinite force applied to
one end. The circles represent passage through an opening,
while the squares represent  motion in the absence of a
membrane. Each data point represents an average over 1000
processes.}
\end{figure}

Figure~\ref{phant_pic} depicts a sequence of ``snapshots'' of a
128-monomer polymer going across the membrane. 
Since the maximal separation between adjacent monomers
is two lattice spacings, the slopes of the curves are limited by 2. 
Note that for $x>0$ the polymer is almost maximally stretched, 
while the $x<0$  configurations resemble the initial random walk state. 
There is also much similarity between the profile of the polymer for $x>0$,
and the the steady--state configurations of a polymer moving
in the absence of the membrane, as depicted in Fig.~\ref{phant_nomemb}
of Appendix~\ref{app_oneD}.

The results of averaging the translocation time (over 1000 realizations)
are indicated by the circles in Fig.~\ref{inf_F_phant}.
The points appear to fall on a straight line in this logarithmic plot,
with the slope of $1.93\pm0.01$ from a least-squares-fit.
There is a slight upwards curvature, and the effective slope varies 
from 1.84 for points with $N\le 128$ to 1.93 for all the points,
indicating potential crossover effects persisting even for $N=512$.
From this data by itself it is difficult to determine the ultimate slope. 
However, we can compare the results with the times required to cross
an imaginary membrane (i.e. unimpeded diffusion). 
This lower bound which was described in the previous section
(and discussed in detail in  Appendix~\ref{app_oneD}), leads to the
mean passage times depicted by the squares in Fig.~\ref{inf_F_phant}. 
The extrapolated slope for this unimpeded motion is indeed  2.00. 
Since the unimpeded crossing times are indeed shorter, 
the asymptotic exponent in the presence of the membrane has to be larger than
or equal to 2.
(Otherwise, for sufficiently large $N$ the curves will intersect 
causing longer times for passage if the membrane is absent.)
We therefore conclude that the translocation of a phantom polymer in 1D 
should asymptotically scale as $N^2$, saturating the bound obtained previously.

\subsubsection{Two--dimensional self--avoiding polymer}

\begin{figure}
\includegraphics[width=8cm]{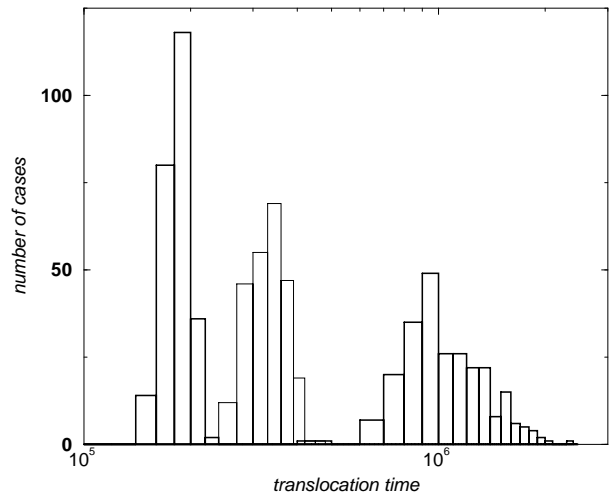}
\caption{\label{timedist_fend}
The distribution of translocation times for a 2D polymer with $N=128$.
Each histogram represents results from 250 independent translocations
for forces $Fa/k_BT=0.25$, 1, and $\infty$ (from right to left) applied to one end.
(The horizontal axis is logarithmic.)}
\end{figure}

We next study the translocation of a self--avoiding polymer in 2D 
as a function of the force $F$ applied to the first monomer. 
Figure~\ref{timedist_fend} depicts the distribution of translocation times 
for a 128--monomer polymer at three values of $f=Fa/k_BT$. 
As $f$ increases from 0.25 to $\infty$, the mean translocation time drops by 
less than one order of magnitude, and the {\em relative width} of the distribution 
decreases somewhat. Note that once $f\gg 1$, the first
monomer always moves in the forward direction, and the results become
independent of $f$.

\begin{figure}
\vskip 1cm
\includegraphics[width=8cm]{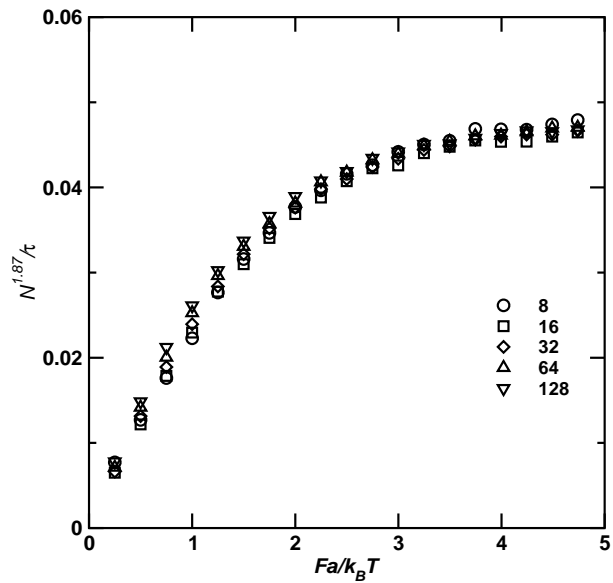}
\caption{\label{NTvsF_end}Scaled inverse translocation time as a
function of the reduced force $Fa/k_BT$ applied to the end--monomer,
for $N=8$, 16, \dots 128.}
\end{figure}

Figure~\ref{NTvsF_end} summarizes the results obtained
for polymer lengths $N$ ranging from 8 to 128, and for a variety of forces. 
Each point corresponds to an average over 1000 realizations. 
The figure depicts the scaled inverse translocation time as a function of 
the dimensionless force $f=Fa/k_BT$. 
The vertical scale has been multiplied by $N^{1.87}$ to produce moderate collapse 
of  data for different $N$s, although as explained further on, we do not believe this
to be the correct scaling factor in the asymptotic regime.
As expected, the curves saturate when $f$ significantly exceeds unity.

\begin{figure}
\includegraphics[width=8cm]{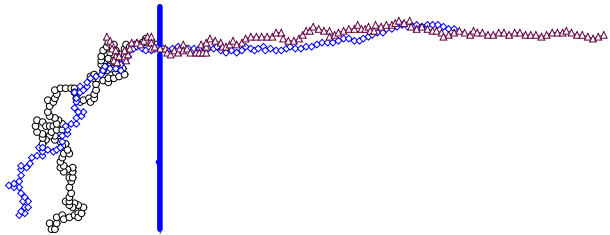}
\caption{\label{pict_fend} Configurations of a polymer of length $N=128$,
pulled through a hole by an infinite force applied to its first monomer. The circles,
diamonds and triangles represent the initial configuration,
and at times $t=60,000$ and 120,000 Monte Carlo time steps.}
\end{figure}

\begin{figure}
\includegraphics[width=8cm]{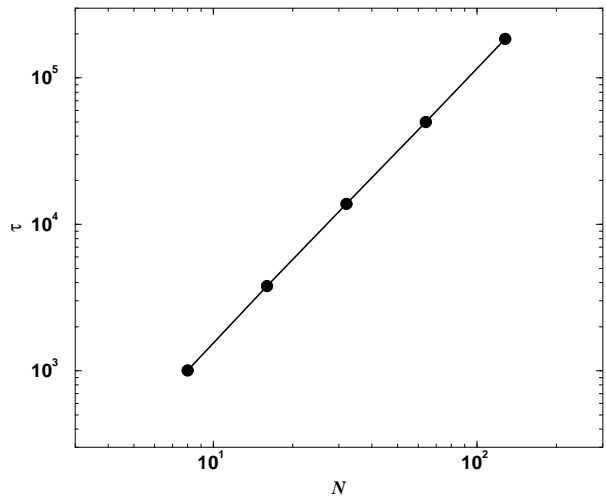}
\caption{\label{infF_end} Logarithmic plot of the dependence
of the translocation time $\tau$ on polymer length $N$, when an
infinite force is applied to the end--monomer. The line
is a fit to a power law dependence with exponent 1.875.}
\end{figure}

Note that all the points in Fig.~\ref{NTvsF_end} belong to the 
regime where $\tilde{f}\gg 1$, i.e. when the shape of the polymer 
is expected to be different from equilibrium, and stretched. 
This is confirmed in Fig.~\ref{pict_fend}, which depicts
configurations of the polymer in the process of translocation under the 
action of an {\em infinite force}. The front end of the polymer is quite 
stretched, somewhat resembling a {\em directed} random walk, suggesting 
that self--avoiding interactions play a secondary role in this limit.
If the front part of the polymer controls the translocation time, 
it should have the same scaling with $N$ as the corresponding 
time for a phantom polymer, i.e. we expect $\tau\propto N^2$.
Figure~\ref{infF_end} depicts the dependence of $\tau$ on $N$;
the effective exponent in this range is $1.875\pm0.005$.
Although smaller than two, it is close to the value of the effective 
exponent of a phantom polymer in this range of lengths $N$.  
Given the bound presented earlier, it is reasonable to expect asymptotic
convergence to this value.
However, as far as we can judge by analogy to phantom polymers, we need 
$N$ to be much larger than 1000 to see an exponent of 2.

\subsection{Chemical potential difference}

When the environments on the two sides of the membrane are 
different, the monomer at the pore experiences a force pushing it
to the more favorable side.
As explained previously, this form of forcing leads to yet a different
form of asymptotic behavior which is once more explored using our
two numerical models.

\subsubsection{One--dimensional phantom polymer}

\begin{figure}
\includegraphics[width=8cm]{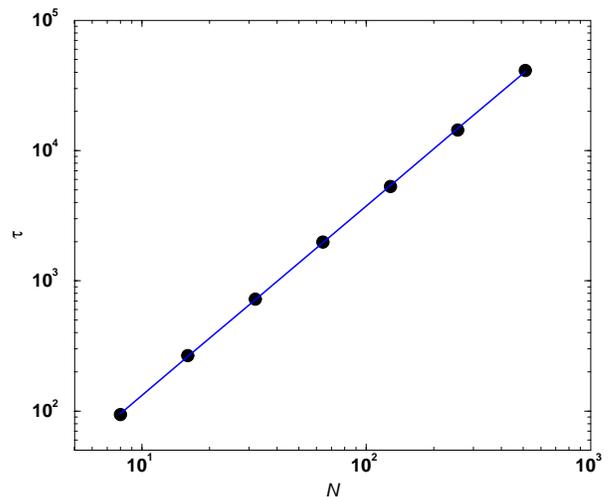}
\caption{\label{inf_ch_phant} Logarithmic plot of the mean
translocation time as a function of polymer length for a
one--dimensional phantom polymer subject to an infinite chemical
potential difference. The solid line is a fit to a power
law with exponent 1.45.}
\end{figure}

Figure~\ref{inf_ch_phant} depicts the $N$--dependence of the
translocation time $\tau$ for a 1D phantom polymer 
under the influence of an infinite potential difference,
i.e. when the monomer at the pore can only move to one side. 
The data on the logarithmic plot are fitted to a straight line
with exponent $1.45\pm0.01$, although there is a slight upward
curvature even for $N=512$. 
Note that in the limit of large chemical potential difference, the
monomers that have already crossed to the right side no longer
play any role in the translocation process, which is thus constrained
by the dynamics of the monomers remaining on the left side. 
In Sec.~\ref{sec_unimpeded} we argued that the translocation
time of an {\em unimpeded} phantom polymer should scale as $N^{3/2}$,
since the leftmost monomer must travel a distance of order $N^{1/2}$
with a velocity of order $1/N$.
Thus 3/2 should be a lower bound for the 
exponent characterizing the scaling of the translocation time with  $N$. 
By comparing this limit with our numerical results, we
conclude 3/2 to be the true asymptotic form describing our simulations. 

\subsubsection{Two--dimensional self--avoiding polymer}

\begin{figure}
\includegraphics[width=8cm]{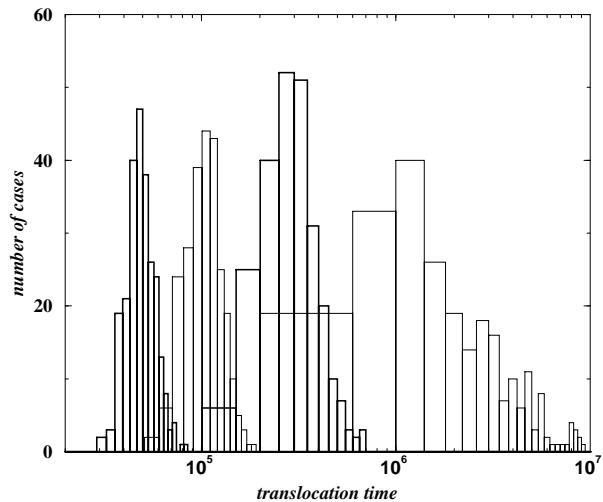}
\caption{\label{timedist_bw_chem} Distribution of translocation
times of a 64--monomer polymer subject to chemical potential differences 
of $\Delta\mu/k_BT =0$, 0.25, 0.75, and 2 (right to left). (The horizontal
axis is logarithmic.)}
\end{figure}

\begin{figure}
\includegraphics[width=8cm]{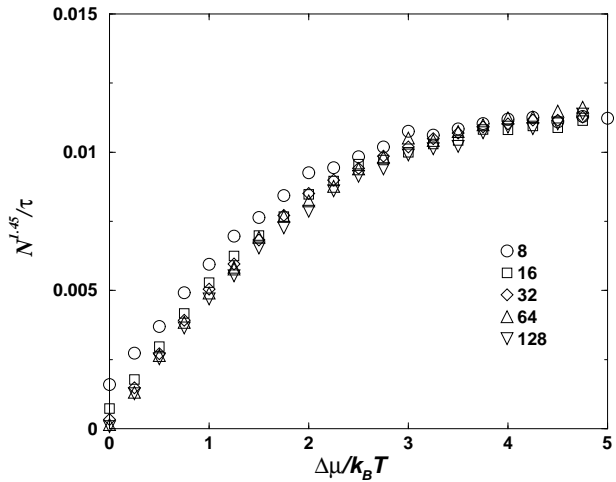}
\caption{\label{CurvsF_chem} Scaled inverse mean translocation
time as a function of reduced chemical potential difference
$\Delta\mu/k_BT$, for $N=8$, 16, 32, 64 and 128.}
\end{figure}

Figure~\ref{timedist_bw_chem} presents the distribution of translocation times
for a polymer with $N=64$ at several values of the dimensionless chemical
potential difference $\Delta\mu/k_BT$.
These distributions are quite wide --- although they become (relatively) 
narrower with increasing $\Delta\mu$, the width of the distribution is of
the same order as the average even in the limit of an infinite $\Delta\mu$.
As expected, the average translocation times decrease and saturate
when $\Delta\mu/k_BT$ exceeds unity. 
The results for different values of $\Delta\mu/k_BT$
and $N$ can be approximately collapsed by scaling $\tau$ with
$N^{1.45}$, as shown in 
Fig.~\ref{CurvsF_chem} (where each point represents
an average over 1000 independent runs). The quality of the collapse
is very poor, and we shall argue that $N^{1.45}$ {\em is not} the
expected asymptotic power.

\begin{figure}
\includegraphics[width=8cm]{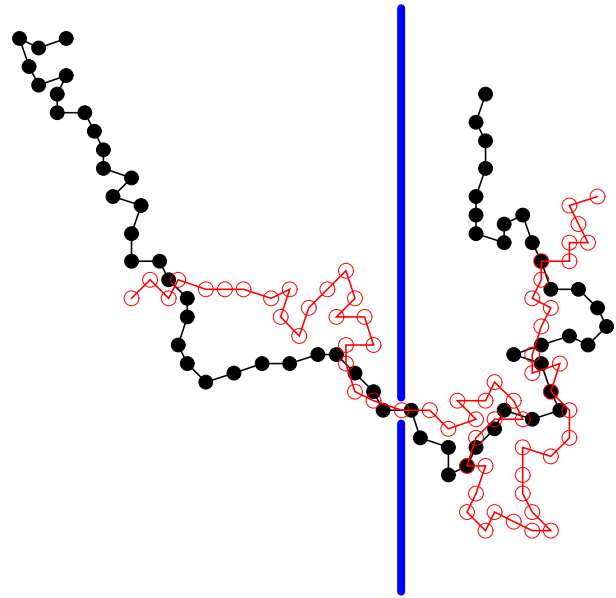}
\caption{\label{transloc_pict_chem} Configurations of a 
polymer of length $N=64$ crossing  a membrane from left to right under
an infinite chemical potential difference. Full and open symbols
represent times of $t=10,000$ and 25,000 Monte Carlo steps, respectively}
\end{figure}

Since (with the exception of $\Delta\mu=0$), the points in Fig.~\ref{CurvsF_chem} 
correspond to a strong force at the pore, we expect the
configurations of the translocating polymer to be different from those of a
polymer in equilibrium.
To explore this difference, in Fig.~\ref{transloc_pict_chem} we show
a pair of configurations for an infinite chemical potential difference.
We see that fast translocation results in a higher density of monomers 
immediately to the right of the pore,  which may in principal slow down the process.
(Recall that in the case of phantom polymers the monomers that have passed
through the hole have no further influence.)
Nevertheless, the whole process should still be bounded by the corresponding
time for passage of an unimpeded polymer, as discussed in Sec.~\ref{sec_unimpeded}.
For considerations of this bound, the relevant time corresponded to motion of the 
leftmost monomer over a distance of size $R\propto N^\nu$, leading to a
time scale growing as $N$ to a power of $1+\nu$ which in 2D is 1.75. 

\begin{figure}
\includegraphics[width=8cm]{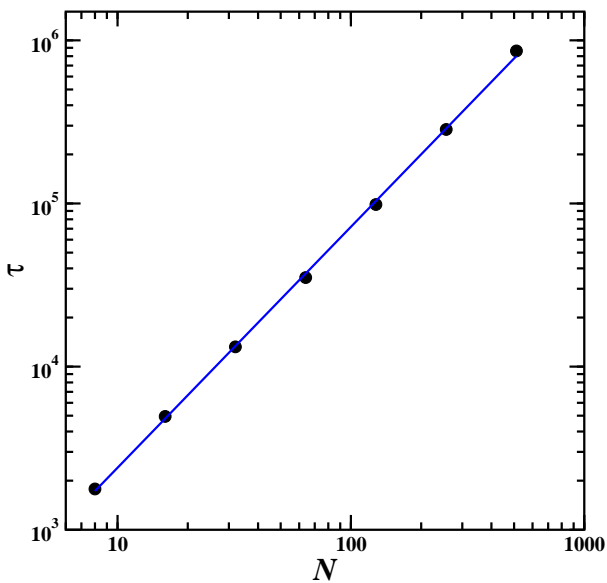}
\caption{\label{InfF_TauvsN_chem} Logarithmic plot of the
dependence of the translocation time $\tau$ on polymer length $N$,
for an infinite chemical potential difference. The solid line
is a fit to a power law dependence with exponent 1.53.}
\end{figure}

The optimal data collapse of the inverse translocation times for
$N\le 128$ leads to an exponent 1.45 (with a rather poor data collapse).
Therefore, we extended our simulations for the case of the
infinite $\Delta\mu$ to larger values of $N$. 
Fig.~\ref{InfF_TauvsN_chem}
represents the dependence of mean translocation time on $N$, for
polymer lengths up to 512. Data points with $N\le 128$ correspond
to averages over 1000 independent simulations, while $N=256$ and
512 include 300 and 130 runs, respectively. The
effective slope of the fit for data points below $N=128$ is
$1.45\pm0.01$, while all the data points produce an effective
slope of $1.53\pm0.01$. Also by directly measuring the effective
slopes between successive pairs of points we definitely see an
increase, with the last pair of points giving a
slope of $1.60\pm0.03$. However, the increase is very slow, and the
uncertainties are too large to enable a reliable extrapolation to
large $N$. By comparing the results to data obtained in the
simulations of phantom polymers we believe that eventually the
exponent will reach 1.75; however, this will probably happen
only for $N$ significantly larger that 1000.

\section{Discussion}

Translocation of a polymer through a pore is intrinsically a many body problem
involving collective and cooperative motion of monomers crossing a membrane.
If the process is sufficiently slow, it is  possible for the segments on the two sides
to come to equilibrium, in which case the dynamics of the translocation
coordinate (the number of monomers on one side) is similar to Brownian
motion of a single particle.
However, the assumption of equilibrium must necessarily break down for long enough
polymers; even more drastically when the polymer is pulled to one side
by a force.
We argued previously \cite{ckk} that the collective motion of the whole polymer
slows down the translocation to the extent that the corresponding dynamics is
anomalous and sub--diffusive.
This is explicitly verified in this paper by quantifying the temporal correlations
of the fluctuating translocation coordinate. We note that there is a general
theoretical framework \cite{metz} for anomalous dynamics of a single variable,
which may be profitably applied to this problem.

One of the objectives of this paper was to find how dynamical anomalies
affect the motion of the polymer under the action of a force.
With this in mind, we also emphasized that the method by which the polymer is
forced is quite important. In particular, pulling the polymer by one end
leads to stretched configurations, and slower overall dynamics, 
compared to applying a chemical potential difference (which can
modify the densities on the two sides).
While pulling the polymer by optical tweezers is not the currently favored
method for artificial translocation of biopolymers, for potential 
applications such
as decoding the sequence, it should offer a better controlled procedure
(whether by itself or in conjunction with a voltage difference).

To understand the time scales involved in forced translocation, we 
initially provided 
what at first glance appear to be a quite loose lower bounds by analogy to 
{\em unimpeded motion} of a polymer,
i.e. neglecting the constraints imposed by passage of the polymer through a
hole in a wall.
We then performed numerical simulations on two model systems: a 1D
phantom polymer, and a self--avoiding polymer in 2D.
Direct interpretation of the numerical results was made difficult by very
large cross--over effects which persist in the length scales of 100--1000 monomers
accessible to numerical study.
Nonetheless, by appealing to the lower bounds found earlier, we concluded
that (rather surprisingly) the actual translocation times scale in the same way
as in the limit of unimpeded motion.
Thus the constraints from the collective motion of the whole polymer turn out
to be at least as important as those imposed by the requirement of 
passage through a hole.

The experiments of Ref.~\cite{meller} suggest that in the range of 10 to 100 base pairs,
the pulling velocity of single-stranded DNA through a nanopore is independent of
$N$, but with a non-linear dependence on the applied force.
In Ref.~\cite{ckk}, we briefly speculated whether such behavior may be consistent
with anomalies associated with polymeric constraints.
As demonstrated in this paper, such constraints result in time scales (and hence
pulling velocities) which must depend on $N$ for large enough $N$.
The only case where we observe a non-linear force-velocity relation which is independent
of $N$ is when short (hence equilibrated) polymers are pulled by a force
applied to one end. 

We hope that these results encourage further experimental and analytical 
studies of forced translocation. 
In particular, it would be interesting to better characterize the manner in which
external forces act on the polymer,
even in the case of a voltage difference across the membrane.
Hydrodynamic effects, not considered in this paper, are also likely to play
an important role.
Some of these effects can be included in more realistic numerical simulations,
although in that case one should keep in mind the rather long crossover times
that appear to be intrinsic to this process.

\begin{acknowledgments}
  This work was supported by the Israel Science
Foundation grant No. 38/02 (Y.K.), and by the National Science
Foundation through grant No. DMR-01-18213 (M.K.).
\end{acknowledgments}

\appendix
\section{The one--dimensional phantom polymer model}\label{app_oneD}

Self--avoiding interactions are integral to
understanding the statics and dynamics of real polymers.
Nevertheless, it is useful to 
study  one--dimensional {\em phantom} polymers,
with no interactions between monomers which are not
adjacent along the chain. Independently of the
details of the interactions of adjacent monomers, long phantom
polymers are {\em harmonic} \cite{deGennes_book}, in the sense that
the probability distribution of the distance $|{\bf r}_i-{\bf
r}_j|$ between monomers $i$ and $j$, approaches a Gaussian for
large $|i-j|$. In
practice, such behavior already appears for $|i-j|\approx 10$, and on
such a ``coarse-grained'' level one can view the polymer as
consisting of monomers connected by springs whose energy is
proportional to $({\bf r}_i-{\bf r}_j)^2$. In this limit,
certain aspects of phantom polymer dynamics can be analyzed
analytically, and we can compare the expected
asymptotic behavior with the numerically observed dynamics. Such
treatment provides both a better view of crossover effects, and
produces some insights into the dynamics of more realistic
(self--avoiding) models.

We employ a phantom polymer model in which
the monomers are restricted to sites  of a 1D lattice with spacing $a$.
The polymer connectivity is implemented by requiring the distance
between adjacent monomers not to exceed two lattice constants. 
An elementary Monte Carlo (MC) step consists of  randomly picking
a monomer and attempting to move it in a randomly selected direction.
If an external force $F$ is applied to the first monomer of
the chain, then the probability to make a step in the direction
opposite to the  force is proportional to $\exp(-aF/k_BT)$. 
A MC time unit corresponds to $N$ attempts to move monomers. 

\begin{figure}
\includegraphics[width=8cm]{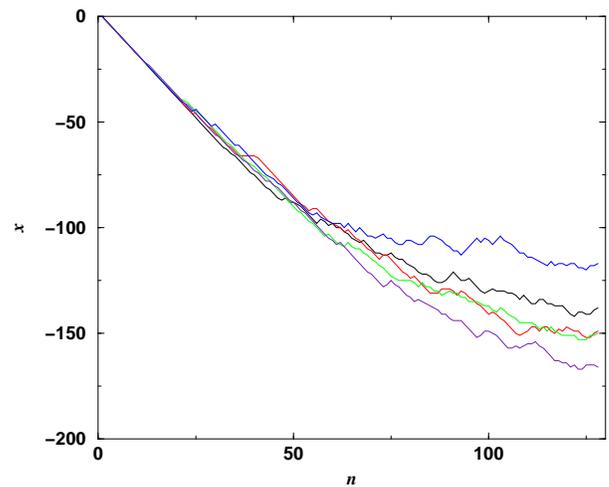}
\caption{\label{phant_nomemb}Five profiles of a
1D phantom polymer of length $N=128$, with an infinite
force applied to one end in  steady state  motion.
The curves are displaced along the vertical axis such that the
position of the first monomer is at $x=0$.
}
\end{figure}

We first performed simulations of polymer motion
when $F\to\infty$, in the absence of a membrane. 
Figure~\ref{phant_nomemb} depicts several examples of spatial
 configurations  in steady state, i.e. when the initial spatial
configuration has been forgotten. 
The resulting profiles can be explained analytically by noting that in
the Gaussian limit the
equation of motion of a 1D phantom polymer is given by
\begin{equation}\label{gauss_chain}
\frac{\partial^2x}{\partial n^2}=\frac{\partial x}{\partial t}\ ,
\end{equation}
where we have used dimensionless units, (distance in lattice constants,
and time in MC units, which leads to a (monomer) diffusion constant
of order one), and omitted prefactors of order unity.
The model used in the MC simulations has a finite stretchability,
while a Gaussian chain can be stretched indefinitely. However,
once the scaled force $f\equiv Fa/k_BT$ becomes
significantly larger than unity, the dynamics 
becomes essentially independent of $F$, as a backwards step of
the end point has negligible probability. Thus, instead of using
$F\to\infty$, we can get set $f=1$, which leads to the boundary condition 
$\partial x/\partial n=-1$ at the beginning of the chain. 
For such a boundary condition we can easily find a stationary solution of
Eq.~(\ref{gauss_chain}), as
\begin{equation}\label{gauss_sol}
x(n)=vt+\frac{v}{2}(n-N)^2 \ ,
\end{equation}
where $v=1/N$. Simulations, of our model polymer in the absence of a
membrane indeed confirm that  the stationary state velocity
is proportional to $1/N$. Moreover, the actual shape of the polymer
in steady state, as depicted in Fig.~\ref{phant_nomemb},
is similar on average to Eq.~(\ref{gauss_sol}).

\begin{figure}
\includegraphics[width=8cm]{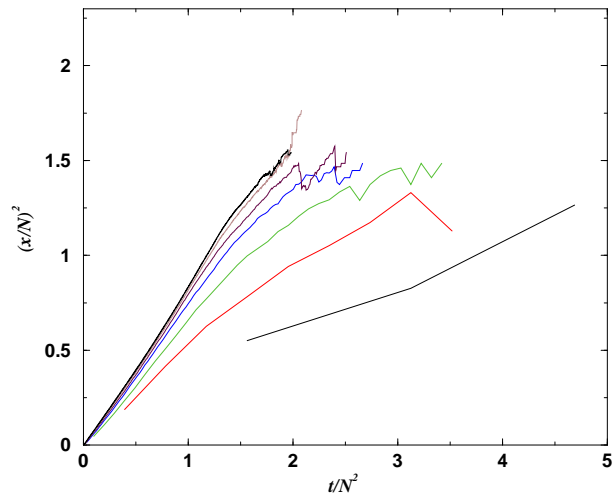}
\caption{\label{imag_wall}
Scaled coordinate $x$ of the first monomer
as a function of scaled time $t$, for $N=8$, 16, \dots, 512 (right to  
left), in the absence of a membrane.}
\end{figure}

The solution in Eq.~(\ref{gauss_chain}) can also be used to understand
the crossover to stationary motion, starting from a relaxed initial state.
It is easy to see that the initial velocity of a point to which the 
force is applied is of
order unity. As time progresses, this velocity decays as
$t^{-1/2}$, until after time $\tau\approx N^2$ it reaches its final
(stationary) value of order $1/N$. The monomer at the opposite end
does not feel the external force in the beginning, and starts 
moving with velocity of order $1/N$ after time $\tau$. 
We also performed simulations of unimpeded polymer motion
mimicking the translocation set-up, by considering an imaginary
membrane located at position $x=0$ which has no effect
on the motion of monomers. 
The initial configuration was chosen by equilibrating the polymer on 
one side ($x<0$), with the first monomer fixed to $x=0$. 
Then an infinite force was applied to the first monomer, and its
position $x$ was tracked as a function of time, until all the 
monomers crossed to $x>0$. 
For every $N$, the results were averaged over 1000 independent runs.
In these simulations the end point only needs to move a distance of order 
$R_g\sim N^{1/2}$ to cross to $x>0$.
Moving at a steady velocity of $1/N$, this would take a time of order
$N^{3/2}$,  which is significantly shorter than $\tau$. 
Thus the time $\sim N^2$ required for the last monomer to start
feeling the force sets the time scale for a polymer
with a large force applied to its first monomer, to move a distance of order
of its radius of gyration. The measured $N$--dependence of the
translocation time is depicted by squares in Fig.~\ref{inf_F_phant}. 
There is a slight curvature in the logarithmic plot and the slope 
approaches $2.00\pm 0.01$, confirming $\tau\sim N^2$.

\begin{figure}
\includegraphics[width=8cm]{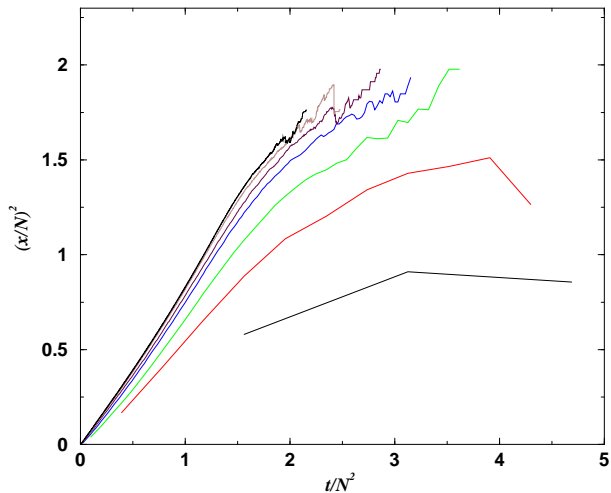}
\caption{\label{true_wall}
Scaled coordinate $x$ of the first monomer of a translocating phantom 1D polymer,
as a function of the scaled time $t$, for $N=8$, 16,\dots, 512 (right to left).}
\end{figure}


The above arguments suggest that $(x/N)^2$ should be a linear function of $t/N^2$.
Thus appropriately scaled plots of the motion of the first coordinates should 
collapse for different values of $N$.
The actual picture appearing in Fig.~\ref{imag_wall} is more complicated: 
while we see an approach to a single curve for the largest $N$, there are very
strong finite size effects for moderate values of $N$. 
The results for crossing the imaginary wall are surprisingly similar to 
those for translocation of the phantom polymer through a hole
(as described in Sec.~\ref{sec_anomal}), as depicted in Fig.~\ref{true_wall}.
(Both plots are obtained by averaging 1000 translocation processes.)  
The linear behavior of these curves close to the origin confirms our expectation 
that at short times the velocity is proportional to $t^{-1/2}$. 
However, for scaled variables around 0.5 the line has a slight curvature, 
which distorts the apparent scaling relations, and creates the illusion of slightly
different exponents.

\medskip

\section{The two--dimensional Self-avoiding polymer model}\label{app_twoD}

We used a 2D lattice fluctuating bond polymer model
\cite{carm} for MC simulations of a self--avoiding polymer.
The monomers are placed on the sites of a square lattice, with the bonds
between adjacent monomers restricted not to exceed $\sqrt{10}$ lattice constants.
The excluded volume between monomers is implemented  by requiring that 
no two monomers can approach closer than 2 lattice constants. 
The membrane with a hole is constructed from a row of 
immobile monomers  arranged in a straight line, with a 3 lattice constant gap 
representing the hole.
Such a hole is small enough to allow only a single monomer to pass through, 
thus enabling a unique identification of the monomer $s$ which
separates the polymer into two segments on different sides of the membrane. 
An elementary MC move consists of randomly selecting a
monomer and attempting to move it onto an adjacent lattice site (in a
randomly selected direction). If the new position 
does not violate the excluded--volume or maximal
bond--length restrictions, the move is performed. 
$N$ elementary moves form one MC time unit.
The first monomer is not allowed to withdraw to the opposite side of 
the membrane. 
Since we are investigating a non--equilibrium process, 
the initial conditions may play an important role. 
We chose an initial state in which the first monomer was fixed
inside the hole, and the remaining polymer was
equilibrated for more than the Rouse
relaxation time \cite{deGennes_book}. 
After such equilibration,  the first monomer was released, and that moment was
designated as $t=0$.

\end{document}